# Neural mechanisms underlying catastrophic failure in human-machine interaction during aerial navigation

**Sameer Saproo[1], Victor Shih[1], David C. Jangraw[2], and Paul Sajda[1]**

[1]Department of Biomedical Engineering, Columbia University, New York

[2]National Institutes of Health, Bethesda, Maryland

E-mail: sameer.saproo@columbia.edu, psajda@columbia.edu

**Abstract.** *Objective.* We investigated the neural correlates of workload buildup in a fine visuomotor task called the boundary avoidance task (BAT). The BAT has been known to induce naturally occurring failures of human-machine coupling in high performance aircraft that can potentially lead to a crash – these failures are termed pilot induced oscillations (PIOs). *Approach.* We recorded EEG and pupillometry data from human subjects engaged in a flight BAT simulated within a virtual 3D environment. *Main results.* We find that workload buildup in a BAT can be successfully decoded from oscillatory features in the electroencephalogram (EEG). Information in delta, theta, alpha, beta, and gamma spectral bands of the EEG all contribute to successful decoding, however gamma band activity with a lateralized somatosensory topography has the highest contribution, while theta band activity with a fronto-central topography has the most robust contribution in terms of real-world usability. We show that the output of the spectral decoder can be used to predict PIO susceptibility. We also find that workload buildup in the task induces pupil dilation, the magnitude of which is significantly correlated with the magnitude of the decoded EEG signals. These results suggest that PIOs may result from the dysregulation of cortical networks such as the locus coeruleus (LC) - anterior cingulate cortex (ACC) circuit. *Significance.* Our findings may generalize to similar control failures in other cases of tight man-machine coupling where gains and latencies in the control system must be inferred and compensated for by the human operators. A closed-loop intervention using neurophysiological decoding of workload buildup that targets the LC-ACC circuit may positively impact operator performance in such situations.

# 1. Introduction

Superior human performance in complex tasks such as piloting a modern jet fighter or driving a Formula 1 car requires goal-directed navigation while operating within dynamic physical constraints or error margins. Such performance requires a careful balancing of cognitive resources, maximizing task engagement while keeping autonomic stress response in check. A failure to maintain this balance can result in catastrophic accidents. For instance, pilot-induced oscillations (PIO) are a dangerous flight characteristic that can spontaneously develop during periods of demanding task performance, e.g., when landing on the deck of a naval aircraft carrier, and can lead to loss of control and airframe damage (Hurt, 1965).

Although the phenomenon of PIOs has been known in the flight community ever since the advent of manned flight, the underlying factors have not been completely understood. PIOs are defined as unstable oscillations in the longitudinal motion of an aircraft that are inadvertently caused by the pilot's own control input. Traditionally, PIOs have been attributed to non-optimal coupling between the pilot and the aircraft. Spontaneous dampened short-period oscillations are normal, but they can become dangerous if the pilot over-compensates small control errors in a way that increases the amplitude of aircraft oscillations to dangerous levels. Moreover, a pilot's unfamiliarity with the 'feel' of the aircraft (e.g., during training or test flights of prototype airplanes) can increase the likelihood of a PIO (Hurt, 1965).

Previous investigations into PIOs have suggested that an aggressive mindset, high pilot workload, and tight error margins could be contributing factors (Gray, 2005, Gray, 2008). In particular, PIOs have been recreated in laboratory conditions using a 'boundary avoidance task' (BAT) paradigm, which entails gradually reducing the permitted margin of error while the pilot is attempting to closely track a complicated flight trajectory (Gray, 2005, Gray, 2008, Dotter, 2007, Warren, 2006). The BAT paradigm is thought to gradually increase a pilot's cognitive workload, arousal, and task engagement, until the cognitive conditions induce a catastrophic control failure, as in a PIO. However, what these cognitive conditions are, and how they mechanistically induce control failures, has not been determined so far.

We present the first neurophysiological study of PIOs, where we investigated neural factors underlying PIOs using a naturalistic 3D BAT paradigm while simultaneously recording EEG and pupillary activity, which was then used to build a predictive classifier that tracked PIO susceptibility. Our data suggest that heightened error monitoring and error sensitivity in the decision-making circuitry of the brain, along with increased arousal, is coincident with a higher probability of PIOs.

## 2. Methods

*2.1. Experimental Design and Stimuli*

We used NEDE (Jangraw et al., 2014), a 3D virtual environment created using Unity game development platform (Unity Technologies, CA), to create a realistic visuomotor task. This task required maneuvering a high-speed virtual aircraft in first-person perspective through a series of equidistant glide boxes that defined a complex undulating trajectory (Figure 1a). The environment was rendered on a 30 inch Apple Cinema HD display (1200 × 800, 60 Hz) that subtended 30 × 23 degrees of visual angle.

The virtual aircraft could be maneuvered in the pitch axis using a flight joystick (Attack 3 Joystick, Logitech S.A.), but with yaw and roll controls disabled. Glide boxes were placed every 500 m in the virtual environment, while the aircraft moved forward with a constant velocity of 250 m/s. Thus, a subject had to navigate through a glide box approximately every 2 seconds. The trajectory formed by glide boxes was modeled as a weighted sum of 3 sinusoids of varying periods and amplitudes in the pitch axis (Figure 1b, Supplemental Figure 1). Glide box boundaries were task critical; failure to navigate through even a single glide box ended the flight abruptly. The task difficulty was manipulated by decreasing the size of the glide boxes at regular intervals (30s) during each trial. Therefore, the user-controlled flight during each trial (max 90s) could be divided into 3 distinct epochs of identical glide box trajectory but with steadily increasing difficulty; this is in consonance with previous 2D BAT investigations (Dotter, 2007). Supplementary Figure 1A shows a second glide path trajectory used in the experiment. Each trial started with 2 seconds of passive fixation (white cross in the middle of a blue screen) followed by 4 seconds of passive flight through the virtual environment towards the first glide box. The subject's joystick input had consequence only after the virtual aircraft had passed through the first glidebox in the trial.

The specific characteristics of the aircraft's response to control input play a critical role in the causation of PIOs, with a more oscillatory pitch response as well as response lag leading to higher chance of a PIO (Dotter, 2007). Therefore, the pitch response of the virtual aircraft in our experiment incorporated both of these elements; an initial lag (0.2s or 0.3s) with a subsequent oscillatory movement of the aircraft in response to a step input of the joystick (Supplemental Figure 1). The maximum instantaneous and sustained pitch response was limited to 40 deg/s and

24 deg/s respectively.

*2.2. Subjects*

A total of 12 healthy human subjects (ages 19-33, all right handed, 4 females) with normal or corrected-to-normal vision participated in the study. Informed consent was obtained from all subjects in accordance with the guidelines of the Institutional Review Board at Columbia University. Subjects were provided with a set of written instructions about the task. They were also familiarized with the virtual aircraft controls by providing least 60-120 minutes of flying practice in the NEDE environment prior to the experiment, either on the same day as the experiment (3 subjects) or the previous day (9 subjects). Data from 3 subjects were discarded from final analysis due to system malfunction during the experiment or poor EEG quality, i.e., inability to get a sufficient number (>1) of neural independent components after artifact rejection (see Data pre-processing section).

*2.3. Data Collection*

During each trial, subjects' joystick input and the position of the aircraft were sampled at 60Hz. In addition to motor behavior, subjects' neurophysiological activity was measured at 2048Hz using an EEG system: Biosemi B.V. ActiveTwo AD-box, 64 Ag-AgCl active electrodes, 10-20 montage. All electrode offsets were below 40 mV at the beginning of the experiment. Gaze position and pupil size were recorded using an EyeLink 1000 eye tracker (SR Research, Ontario, Canada) at a 1000 Hz sampling frequency. The subject's head was stabilized during data collection using a chin and forehead rest. A 9-point calibration of the eye tracker was performed before each block of trials, and the subjects were instructed to not move their head between or during trials (they were provided regular breaks however). Note: stimulus delivery and behavior recording, EEG recording, and eye tracking were performed on 3 different computers simultaneously, with the respective data synced post-hoc (Jangraw et al., 2014).

A total of 32, 40, or 48 flight runs were observed per subject, with the total experiment time not exceeding 1 hr. Note that given 2 different pitch response delays and 2 different glide-path trajectories, each unique combination of delay and trajectory was run an equal number of times (e.g., 10 per combination, for a total of 40 trials per subject). This was done to maximize the probability of sampling pilot behavior relevant to PIOs given inter-subject variability in innate performance on the task. Furthermore, slight changes in control parameters and flight trajectory across trials ensured that subjects had minimal learning or 'muscle memory' of the specific experimental parameters, and thus no steady increase in performance throughout the experiment.

*2.4. Data Pre-Processing*

All EEG and pupillometry data were analyzed using the EEGLAB toolbox (Delorme and Makeig, 2004) in MATLAB (The MathWorks Inc., Natick, MA). The recorded EEG signals were first re-referenced to the average of all electrodes and then band-pass filtered to 0.5-100Hz using a Hamming windowed FIR filter. The result was then notch filtered at 60Hz to remove line noise, and finally down-sampled to 256 Hz. This pre-processing pipeline produced 'raw' datasets that contained signals from neural, ocular, and muscular sources, as well as non-physiological artifacts.

To isolate the purely neural component of the EEG data, we used the following procedure: we first reduced the dimensionality of the EEG data by reconstituting the data using only the top 20 principal components derived from Principal Component Analysis (PCA). Thereafter, an Independent Component Analysis (ICA) decomposition of the data was performed using the Infomax algorithm (Bell and Sejnowski, 1995). We then used an ICA-based artifact removal algorithm called MARA (Winkler et al., 2011) to remove ICs attributed to blinks, horizontal eye movements (HEOG), muscular activity (EMG), and any loose or highly noisy electrodes. MARA performs automatic IC classification using a linear classifier trained on time-series, spectrum, and scalp map features of a large dataset of labeled IC artifacts. MARA assigns each IC a probability of being an artifact; we removed components with probabilities above 0.5.

*2.5. EEG Data Classification*

The 64-channel EEG signals recorded during each trial were split into 1500 ms epochs that were centered at the onset of each stick movement. Thus, each epoch was construed as a unique data point for classification. We used spectral power as the classification feature; therefore, a spectrogram of the entire continuous data was computed using a sliding Short Time Fourier Transform (STFT) with a 128 sample Hamming window and 64 samples of overlap between windows, yielding five 500 ms windows of frequency data for each electrode. For further analysis, frequencies were separated into bands of interest: delta band consisted of frequencies 1-3 Hz, theta band 4-7 Hz, alpha band 8-15 Hz, beta band 16-31 Hz, and gamma band 32-55 Hz. Spectral information from 56-128 Hz was not used in the classifier.

Classification was performed either using the spectral power in different frequency bands as features (delta only, theta only, alpha only, beta only, gamma only) or using the power from all frequencies (all bands, Figure 3). Each data point was given a class label according to the nearest boundary size at the time of stick movement during the trial. In one classification regime (which we call LvsMS), the first class includes all stick movements made during navigation under large boundaries, and the second class includes all stick movements made during navigation under medium and small boundaries. In another classification regime (time-on-task or ToT), data points generated only within large boundaries were sub-divided according to their temporal occurrence into classes (first half or second half). Finally, the MvS classifier used the weight vector learned from the LvMS regime to classify Medium vs Small classes; this was done in order to show that the workload component (weight vector) was generalizable to different absolute boundary sizes. We used N-fold cross-validation to generate results, usually with both training and test data derived from the same subject, except in one case where leave-one-subject-out cross-validation was used to investigate whether the PIO classifier was generalizable to novel subjects.

*2.5.1. Classification Algorithm*

The high dimensionality of data – there are 17,600 features per data point when using all frequency bands (1-55Hz) – required the use of a recently developed algorithm FaSTGLZ (Conroy et al., 2013), for efficient linear classification. FaSTGLZ classifies input data $x \in \mathbb{R}^D$ with binary class labels $y \in \{0, +1\}$ by using logistic regression to create a separating hyperplane in the feature space that is parameterized by a normal vector $w = (w_1, \dots, w_D) \in \mathbb{R}^D$. For the sake of simplicity, the classifier bias is estimated by incorporating a constant $x_{D+1} = 1$ and bias term $w_{D+1}$ into the classification.

The posterior probability of the class label $y_i$ for each data point $x_i$ is modeled as a sigmoid function:

$$p(y_i = 1|x_i, w) = \frac{1}{1 + \exp(-x_i^T w)} \quad (1)$$

For logistic regression, denoting $p(y = 1|x, w) = \mu(x^T w)$, the negative log-likelihood is given by

$$\mathcal{L}(w) = -\sum_{i=1}^{N} y_i \log\left(\mu(x_i^T w)\right) + (1 - y_i)\log\left(1 - \mu(x_i^T w)\right) \quad (2)$$

A common problem with Maximum-likelihood estimators is the severe over-fitting of high dimensional training data. FaSTGLZ mitigates such over-fitting by using a penalized likelihood method based on $L_2$-regularization that seeks to minimize:

$$J(w) = \mathcal{L}(w) + \lambda w^T L w \quad (3)$$

If the norm $L$ is any symmetric positive semi-definite matrix and $\lambda$ a real-valued scalar, then $J(w)$ would be a convex function, which is optimized by FaSTGLZ using the 'Alternating Direction Method of Multipliers' (ADMM) procedure. ADMM uses variable splitting to divide the main optimization into two simpler sub-procedures to minimize a differentiable objective and to solve a soft-thresholding operation. This allows the simultaneous training of high dimensional models across bootstraps, cross-validation folds, and permutation tests, thus considerably speeding up classifier learning. Note that all classifiers in our analyses were learned using 5-fold cross validation with 100 bootstraps for each fold. The optimal lambda values were chosen using a parameter sweep of 100 lambda values between 1e5 and 1e-5; the value that yielded the highest AUC was used for further analysis.

The resulting classifier assigns a set of weights to the feature space used to train the model, such that each multidimensional data-point is projected onto a scalar dimension where the two classes are maximally separated. The classifier features – spectral power of EEG signals – were z-scored across epochs before classification, and therefore the learned classifier weights can be interpreted as the normalized contribution of each frequency at each electrode to the discriminating hyper plane. A positive weight would imply that the classification feature is more correlated with low pilot workload (Larger boundary size) and a negative weight would imply a stronger correlation with higher pilot workload (smaller boundary size), therefore describing the direction of the change in spectral magnitude across boundary size. Furthermore, the entire set of classifier weights (frequency band × time point × electrode) can be localized on the scalp, thus showing the spatial and temporal signature of neural correlates of workload (Figure 3).

## 3. Results

The behavioral data show that our experimental paradigm elicited piloting behavior relevant to PIOs, i.e. there is an increase in the magnitude of PIO features with decreasing boundary size during the boundary avoidance task (Figure 2). Specifically, a reduction in boundary size led to quicker task failure (One-way repeated-measured ANOVA; $F(2,16)=185.3$, $p<0.001$), increased magnitude of joystick force ($F(2,16)=63.59$, $p<0.001$), increased frequency of joystick input ($F(2,16)=22.24$, $p<0.001$), and a rapid increase in the phase divergence between the input and the response ($F(2,16)=25.37$, $p<0.001$). Here, the phase divergence, between the input and the response, was computed by taking absolute the difference between the unwrapped phases of the Hilbert transform of accumulated joystick input and current aircraft heading.

We used the spectral power of stick-locked EEG signals (1500ms around each stick movement) in different canonical frequency bands (delta, theta, alpha, beta, and gamma) as features to classify BAT-induced workload (Classes: Large vs Medium/Small boundaries), and we observed above-chance classification accuracy (chance AUC=0.5) for all subjects (Figure 3A,

Table 1). The choice of the size of the epoch was dictated by the asymptote of classification accuracy across different epoch sizes (Supplemental Figure 2). In order to dissociate the contribution of different frequency bands to workload classification, we computed the classification accuracy for each band separately by filtering EEG signals to the respective band before classification (see Methods for specific frequencies for each band). The results show that the gamma band is the most informative band for classifying BAT-induced workload, as the gamma band classifier approximates the accuracy of a full spectrum classifier. This difference in classification accuracy does not seem to be a consequence of the higher dimensionality of the gamma band, as qualitatively similar results are produced when classifying using the average power in each band (Supplemental Figure 3). We also find that regularly sampling EEG signals (every 2s) for classification did not have a significant qualitative difference from stick-locked EEG signals (Supplemental Figure 4).

The scalp topology of the classifier weights for a full-spectrum classifier suggests that the contribution of delta, theta, and gamma band activity to workload classification is spatially localized. Delta and theta band based classifiers had significant contributions from fronto-central sites, while gamma band modulation had a predominately lateralized somatosensory topography (Figure 3B). Furthermore, the scalp topology for all frequency bands was robust to the variation in the temporal overlap of windows used to compute spectral power using a Fast Fourier Transform (see Supplemental Figure 5 for a classifier with higher temporal overlap).

We tested the robustness and the generalizability of classifiers from individual bands. We found that even though the theta band classifier does not produce high classification accuracy compared to the gamma band classifier, fronto-central theta activity is a more robust signature of workload since MARA-based artifact cleaning of raw EEG affected theta-band classifier the least (Figure 4). These data suggest that fronto-central theta activity might prove to be the best indicator of workload in an operational scenario (i.e., while flying a real fighter plane), due to the significant contamination of EEG signals with potentials from muscle activity. Similarly, training the classifier with data from non-test subjects (hold-one-subject-out cross-validation) impacted the accuracy of theta band classifier the least, further attesting to its generalizability (Figure 5).

Workload can build up due to sustained focal attention required by the task, which would be unrelated to boundaries in the BAT scenario. However, we find that the contribution of time-on-task component to classifier performance for our data was not enough to explain the steadily increasing EEG signatures of workload in our BAT paradigm (Figure 6, Supplemental Figure 6). Furthermore, LvMS classifier (classes: large vs medium/small; data from medium and small boundaries were collapsed into a single class) could also reliably distinguish Medium from Small boundary conditions (MvS). This suggests that our assessed neural correlates (vector of weights normal to the classifying hyperplane) are independent of absolute boundaries and can reliably predict a continuum of workload states. More importantly, data suggest that neural correlates derived from laboratory BAT experiments can be effectively used to provide continuous feedback about pilot workload and PIO tendency in real-time (Figure 7).

Although, we trained classifiers to discriminate EEG signals based on boundary size, we demonstrate that the classifier output can also track PIO tendency. We estimated PIO tendency by creating a metric; the amplitude of Hilbert transform of band-passed stick movement (0.3-1.8Hz, range that is typical of PIOs; (TIAN, 2006, Rzucidło, 2007). We then separated the trials into 4 bands of increasing PIO tendency according to the magnitude of PIO measure in the last 5 seconds of a trial, and compared highest to lowest band (Q1: band lower than $1^{st}$ quartile; Q3: band higher than $3^{rd}$ quartile). We find that there is a significant difference between time-averaged (~10s) classifier output that leads up to a PIO event towards the end of the trial (Q3), compared to trials that did not end in a PIO (Q1), with the temporal trend showing steady divergence (Figure 8).

In addition to EEG, we collected pupillometry data while the subjects performed the experimental task. Data show that subject's pupils dilate with a decrease in boundary size (Figure 9A), directly implicating mental load and arousal (Murphy et al., 2011). We find a significant correlation between EEG-derived classifier output and pupil size for a full spectrum classifier (Figure 9B, one sample t-test; $t(8) = -2.41$, $p = 0.043$). More importantly, we find a significant increase in the correlation between the output of theta band classifier – with fronto-central topography – and the pupil size with a decrease in boundary size during BAT (Large vs. Medium boundaries), suggesting a close anterior cingulate cortex – locus coeruleus norepinephrine (ACC-LC-NE) interaction during induced workload buildup (One-way repeated-measured ANOVA; $F(1,8)= 8.08$, $p=0.022$). This change in correlation was not observed with the full-spectrum classifier ($F(1,8)= 0.02$, $p=0.88$).

**4. Discussion**

We performed the first neurophysiological investigation into the phenomenon of pilot induced oscillations (PIOs), using a boundary avoidance task (BAT) in a gaming/virtual reality environment. We find that our task is able to gradually induce cognitive workload that in some cases causes PIO-like behavior. Furthermore, we find robust EEG signatures of workload in different spatio-spectral bands, with fronto-central theta band the most robust, in terms of being differentiable from potential artifacts. We also find a significant correlation between this EEG activity and pupil dilation due to BAT induced workload. Below we discuss these results within the context of specific circuits in the brain and also possible broader implications of our findings.

The encoding and regulation of error monitoring has typically been associated with the anterior cingulate cortex (ACC), which is believed to be a key brain area for cognitive control and storing of predictive models of our environment (Tervo et al., 2014). The region ACC is believed to be at least partially modulated by the locus coeruleus, a tiny nucleus in the dorsal pons, regulating arousal levels in the brain via the neurotransmitter norepinephrine (Aston-Jones and Cohen, 2005). The link between arousal state and task performance has been shown to be non-linear (Aston-Jones and Cohen, 2005, Gompf et al., 2010). For example, the Yerkes-Dodson curve posits that a mid-level arousal state is the most optimal for task performance, though this "mid-level" is highly task and context dependent. Therefore, the LC-ACC circuit is of particular interest in decision-making under dynamic constraints (e.g., flying an aircraft or driving a vehicle) since dynamically adapting motor control strategies based on assessment of current performance and upcoming task constraints is often key to optimal performance.

Recent work in animal models has shown a tight coupling between the LC- norepinephrine system (LC-NE) and the ACC when animals must dynamically switch between task-based models (Tervo et al., 2014). Specifically, the rats in the experiment faced a computer opponent in a competitive virtual task, where the computer was programmed to counter-predict a rat's behavior. When the LC-NE input to the ACC increased, the rats were less adept at incorporating environmental feedback into their internal model of choice and prediction. However, when the LC-NE input to ACC was suppressed, the rats were able to utilize feedback from the environment more effectively and therefore better model the computer's counter prediction to increase their performance and reward.

Though we cannot directly measure LC-NE activity with scalp EEG, several studies have shown that pupil dilation can be used as a proxy for activity in the LC and thus provides some information of the state of arousal of an individual (Gompf et al., 2010, Joshi et al., 2016). ACC, on the other hand, is more accessible via EEG, with fronto-central theta activity having been identified as a correlate of ACC activation (Cavanagh and Frank, 2014). Thus, by linking EEG activity with pupillary measures, one can potentially, non-invasively, infer the dynamics of the LC-ACC circuit during a complex and dynamic task.

Our results, therefore, can be interpreted within the context of the aforementioned study of LC-ACC interaction (Tervo et al., 2014), and may provide a mechanistic explanation for PIOs. An increase in observed fronto-central theta band power, which in turn is correlated with pupil dilation in our study, could be an indication of the subject switching into a behavioral model associated with high workload state. This might suggest that in a cognitive state associated with high workload, there is an increase in LC-NE input to ACC, which might lead to the subjects sticking with their current internal model of aircraft control, even when the boundaries have changed. This sub-optimal control model might lead to PIOs in certain instances. In contrast, a better strategy would be to incorporate environmental feedback and switch to a different internal model of aircraft control that is better adapted to steering within narrow boundaries.

This interpretation of our results provides an interesting possibility for mitigating PIOs: since the LC-NE system is associated with arousal, using feedback from a hybrid BCI system (hBCI) to dynamically adjust arousal levels may regulate LC-NE input to the ACC, allowing updates to the internal model of the pilot based on the environmental feedback. For example, a hBCI that integrates pupillometry and EEG features could predict when a pilot is entering a state that will likely generate a PIO (as in Fig. 8, red curve), whereupon feedback in the form of a continuous auditory stimulus with calming influence, could be delivered to reduce arousal level and thus reduce LC-NE input to ACC. We hypothesize that an optimally calibrated feedback loop would help regulate LC-ACC interaction, resulting in piloting behavior improvements.

Beyond cases of vehicular control, there is a large class of electronic games, such as the highly popular "Flappy Bird", that resemble a boundary avoidance task; the player controls a character moving at constant speed, avoiding obstacles and boundaries that become tighter as the game progresses. The objective in these games is to go as far along the course as possible. These games are known to be highly addictive, with the gamers repeatedly replaying the course from the beginning after a failure, trying to increase the distance they come along a course before failure. Although a recent attempt has been made to quantify the optimal parameters for such games such that they remain highly playable (Isaksen, 2015), the cognitive factors underlying their addictive nature remains unknown. Our results from the BAT investigation suggests that not only does arousal level (as evidenced by pupil dilation) increase progressively as the boundaries decrease and therefore difficulty increases, but there is a dramatic increase in the cognitive workload due to task monitoring (as evidenced by theta band activity over fronto-central sites). This presents a peculiar hypothesis: perhaps the addictive nature of such games comes from the ability to progressively achieve higher arousal levels, as the subjects improve their ability to increase the ACC-LC coupling to move to the more optimal position along the Yerkes-Dodson curve. Indeed, our data show that the correlation between fronto-central theta band activity (stand in for ACC) and pupil dilation (stand in for LC activity) increases during the course of the experiment.

Finally, our results have relevance beyond the world of tracking physical boundaries as in gaming or vehicular navigation; humans frequently engage in sustained perception-decision-action loops that involve goal and error tracking under dynamic constraints. For example, a financial portfolio manager has to track changing market conditions and reallocate stocks so as to maximize portfolio value while managing risk within prescribed boundaries. Project managers must regularly track project progress and deal with exigencies so as to ensure high quality of work while avoiding unacceptable delays in completion. Viewed generally, these examples are illustrative of rapid decision-making that involves tracking optimal performance while avoiding frequently changing 'failure' boundaries. As with top-gun pilots, Formula 1 champions, top fund managers, and top project managers, the burning question is "what neural markers differentiate stellar performance (and performers) from catastrophic failures under challenging conditions"? Our results suggest that the key insight may lay in the interaction of neural circuitry that is engaged in error monitoring, decision-making, and regulating arousal.

## TABLES

|  | All Bands | Delta | Theta | Alpha | Beta | Gamma |
|---|---|---|---|---|---|---|
| **Stick Locked Classifier** | 0.75±0.02 | 0.56±0.01 | 0.57±0.02 | 0.59±0.02 | 0.65±0.03 | 0.73±0.02 |
| **Regularly Sampled Classifier** | 0.75±0.02 | 0.57+0.01 | 0.58+0.02 | 0.61±0.02 | 0.66±0.03 | 0.72±0.03 |

**Table 1**: Mean area-under-curve (AUC) values for workload classifiers (Classes: Large vs Medium/Small boundaries) using different spectral content (mean±SEM).

**Acknowledgments**

The authors would like to thank Eric Pohlmeyer and Mark Chevillet for their helpful comments and suggestions, and acknowledge Sona Roy and Wha-Yin Hsu for their assistance with data collection. The work was funded by the Defense Advanced Research Projects Agency (DARPA) and Army Research Office (ARO) under grant number W911NF-11-1-0219, the Army Research Laboratory under Cooperative agreement number W911NF-10-2-0022 and the U.K. Economic and Social Research Council under grant number ES/L012995/1. The views and conclusions contained in this document are those of the authors and should not be interpreted as representing the official policies, either expressed or implied, of the US Government. The US Government is authorized to reproduce and distribute reprints for Government purposes notwithstanding any copyright notation herein.

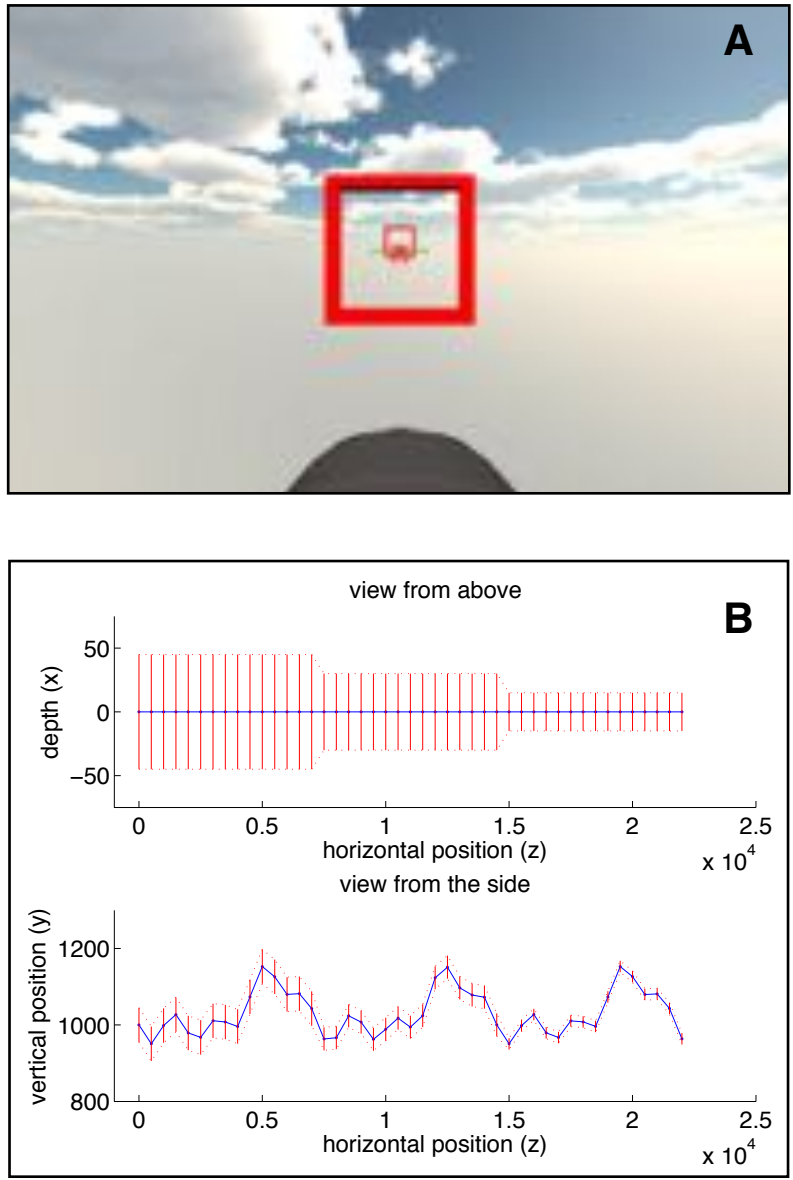

**Figure 1. A)** Screenshot of subject's view during 3D Boundary Avoidance Task (BAT) experiment. Red squares depict waypoint boundaries, while the dotted horizontal green line in the center shows the current heading of the virtual aircraft. **B)** Full flight trajectory with the position and the size of the glide boxes; solid blue line shows the mean path through the center of glide boxes, while solid red lines denote glide boxes. All dimensions are in meters. Virtual aircraft moved steadily in z-axis @ 250 m/s and could be controlled in y (pitch) axis via joystick input. See supplemental figure 1 for the other glide path trajectory used in the experiment.

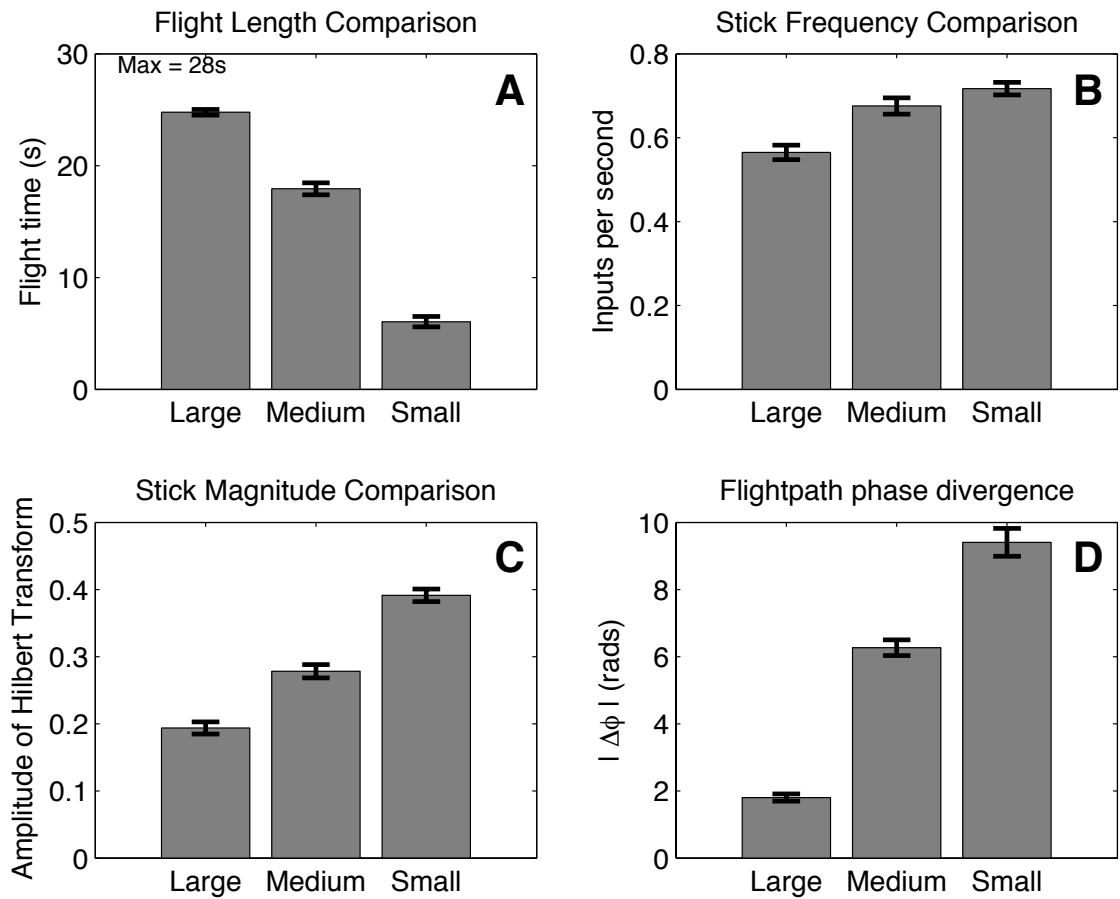

**Figure 2.** Four measures that demonstrate that the BAT paradigm elicited control behavior typical of PIOs; **A)** a reduction in boundary size led to a decrease in flight length before failure (missing a glide box), **B)** more frequent and **C)** larger joystick inputs, and **D)** a quicker increase in the phase divergence between the Hilbert transform of cumulative control input and current aircraft trajectory. Error-bars reflect mean ± SEM across subjects (N=9).

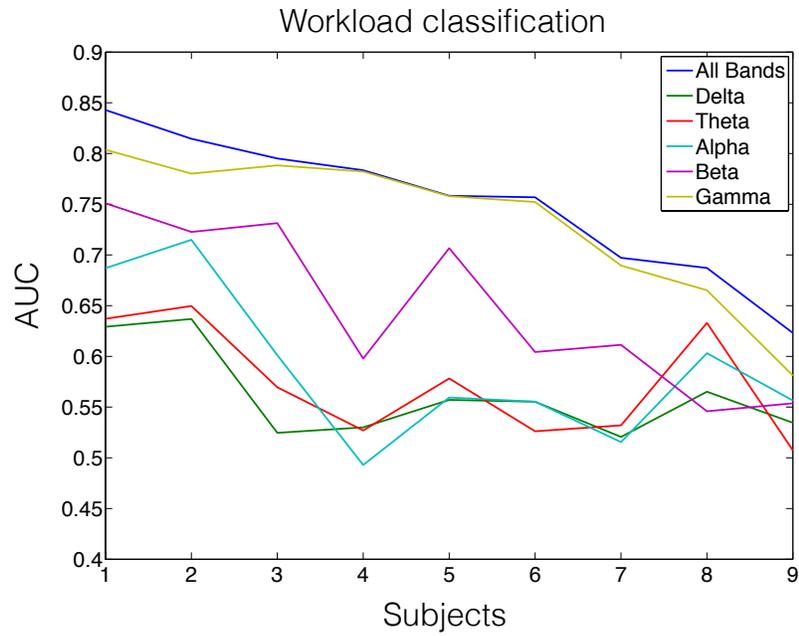

A

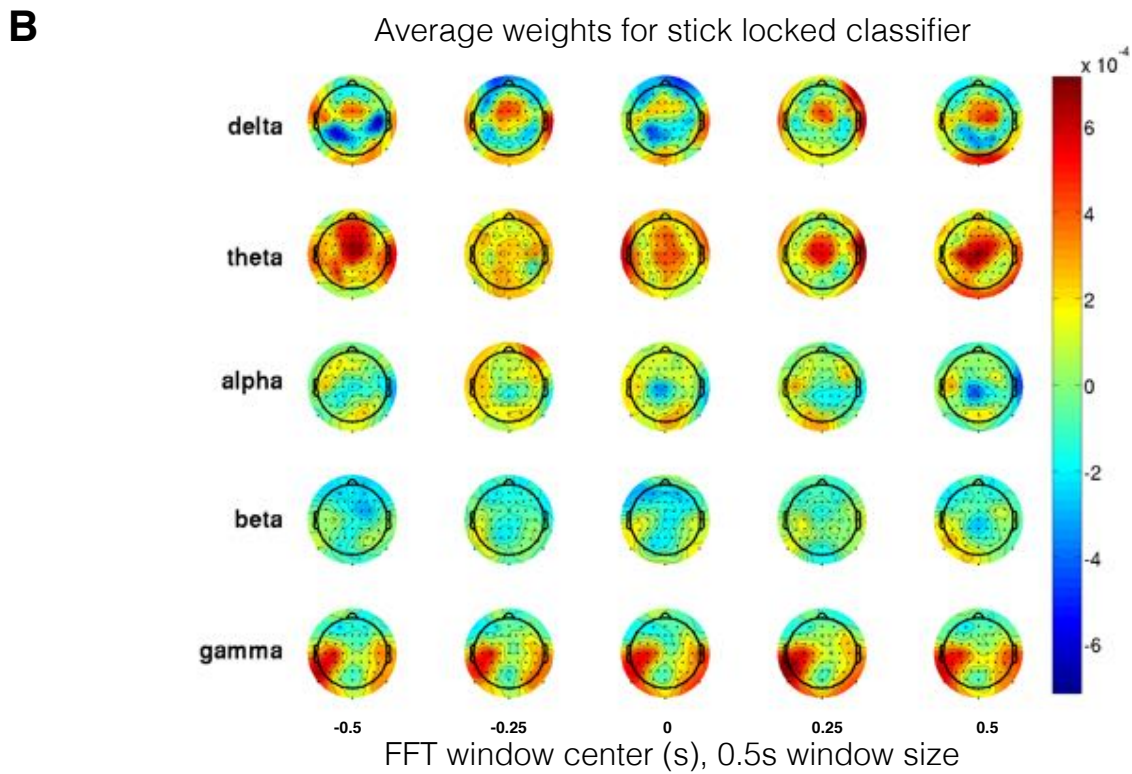

B

**Figure 3. A)** Area under the receiver operating characteristic curve (AUC) for all subjects (in descending order of 'All Bands' classifier AUC), when using information from all spectral bands for classification ('All Bands'), as well as when using only individual bands (delta, theta, alpha, beta, or gamma). Classification was performed using 64-channel MARA-cleaned EEG signals —1.5s epochs around each joystick movement — that were labeled according to the size of the nearest glide path boundary at the time of their generation. **B)** Subject-averaged scalp distribution of normalized weights for the classifying hyperplane that best separated Large boundary from Medium and Small boundaries (scalp map corresponds to 'All Bands' classifier in panel A). Delta and theta band activity from fronto-central electrodes, as well as significant gamma band activity from a lateralized somatosensory topography, seems most indicative of higher workload induced by smaller boundaries.

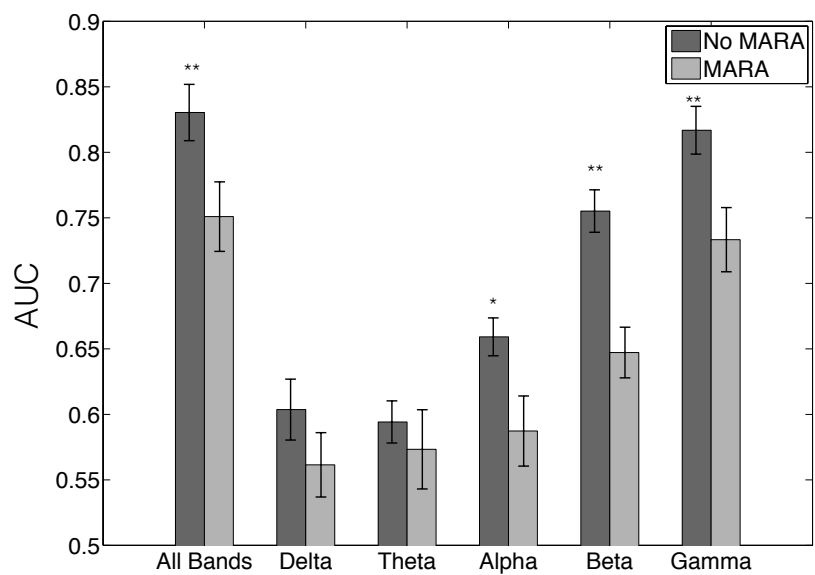

**Figure 4.** Effect of artifact removal, using MARA algorithm, on classification accuracy (Large vs Medium/Small) when using information all bands or when using individual bands. MARA algorithm classifies ICA components in the data as artifact based on a pre-existing labeled set of artifactual ICs, including those for eye movements, muscle movements, and noisy electrodes. Error-bars reflect mean ± SEM across subjects. Paired t-test; * $p<0.01$, ** $p<0.001$. Figures 3, 5-9 show results from MARA cleaned data.

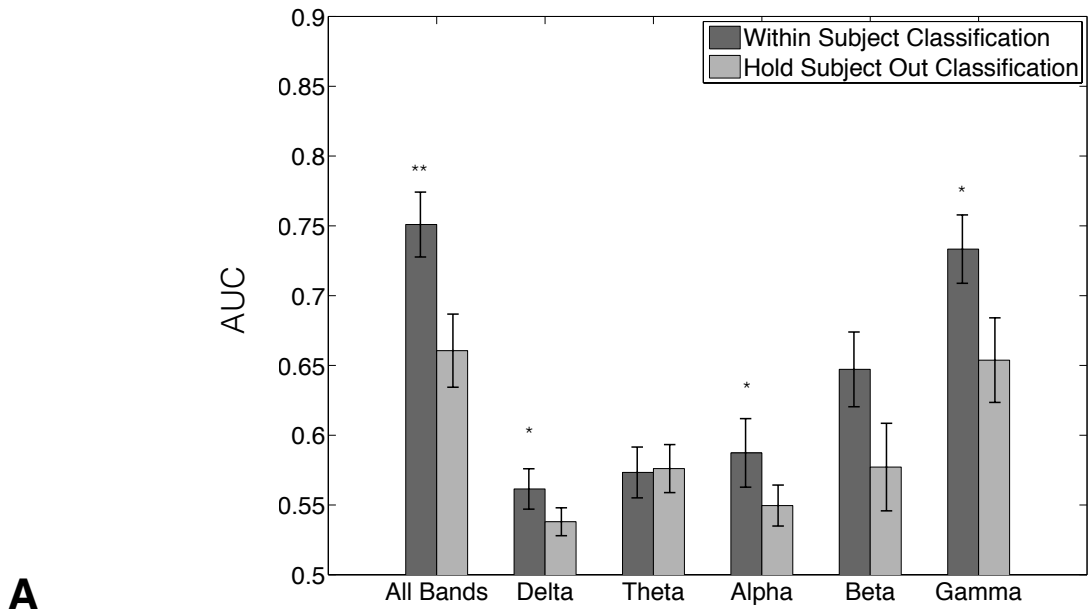

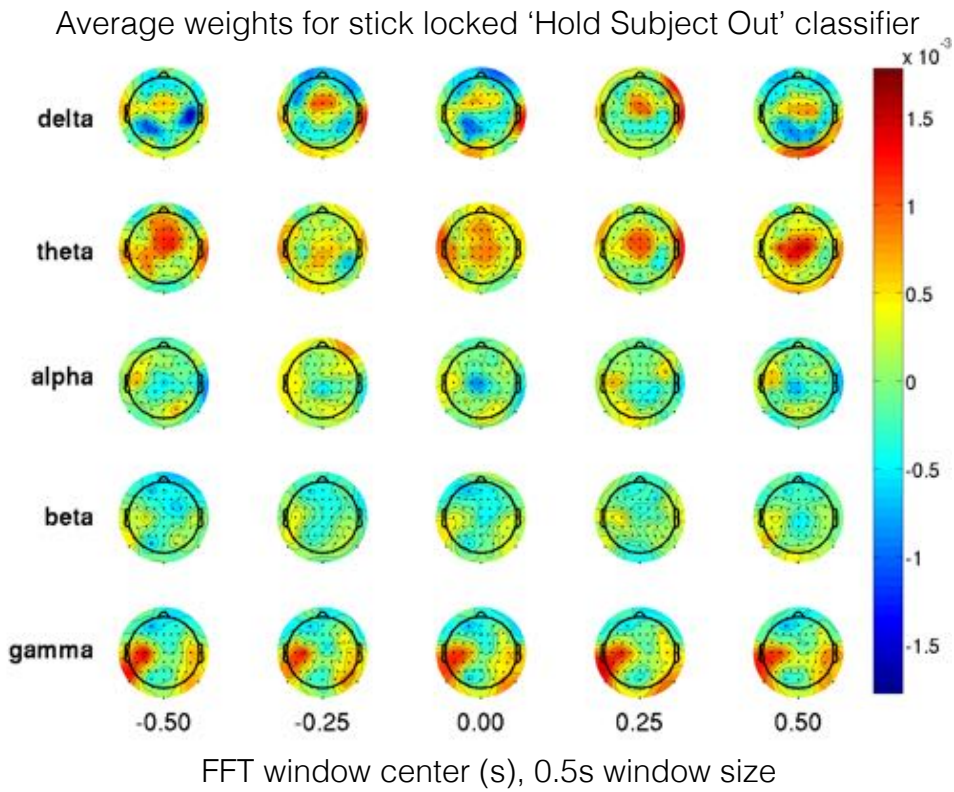

**Figure 5. A)** Comparison of classification accuracy (Large vs Medium/Small) when using 'Within subject' k-fold cross-validation or when using 'Hold Subject Out' cross-validation. Error-bars reflect mean ± SEM across subjects. Paired t-test; * p<0.05, ** p<0.01  **B)** Subject-averaged scalp distribution of normalized weights for the classifying hyperplane that best separated Large boundary from Medium and Small boundaries for 'Hold Subject Out' classification (scalp map corresponds to 'All Bands' classifier in panel A).

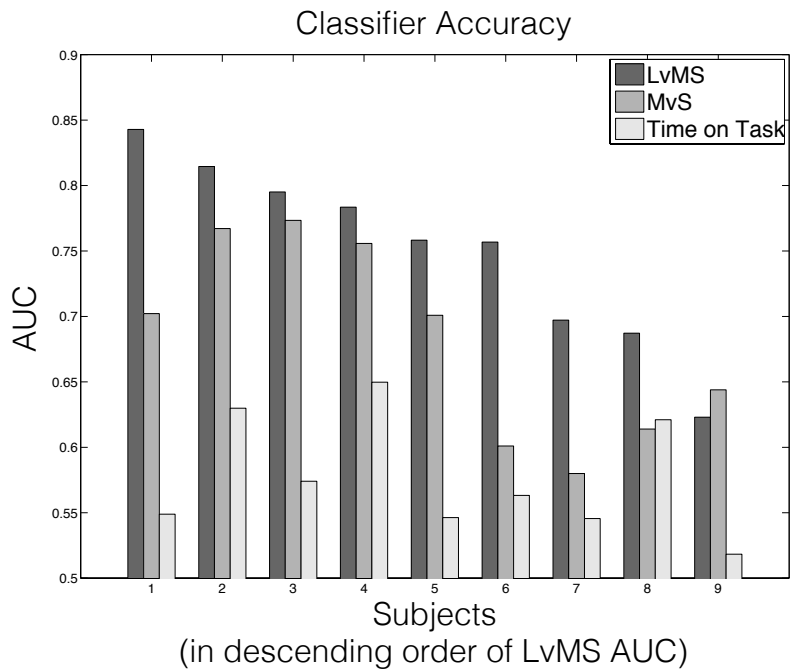

**Figure 6.** Subject-wise Classification accuracy: LvMS, where data from Medium and Small boundaries was combined into a single training class, and the classifier learned to distinguish Large from Medium/Small boundary classes. MvS, used the weights learned from LvMS classifier and classified Medium vs Small boundary classes. Time on Task, when classifying data from the Large boundary class that is labeled according to temporal occurrence - first half or second half of flight within large boundaries.

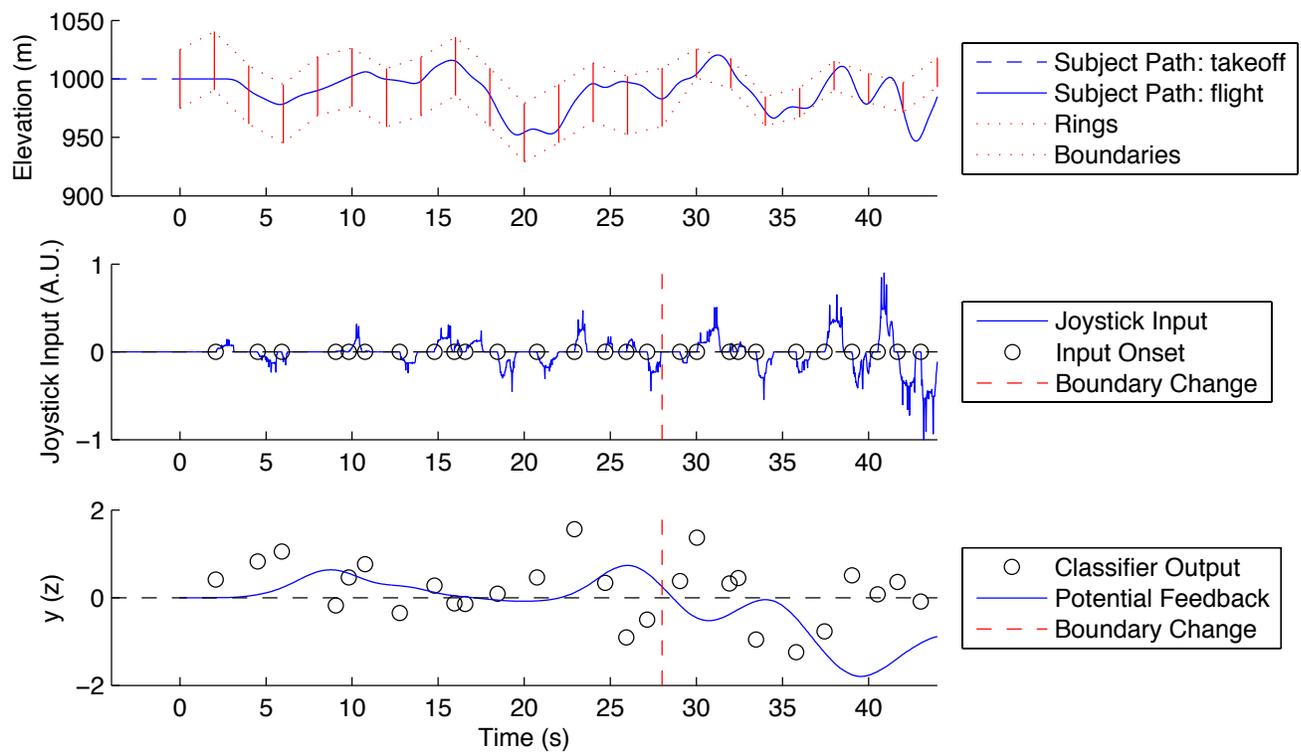

**Figure 7:** A representative trial flight showing the measured behavioral and neural markers across time. Top to bottom: Flight path, control stick movements, commanded and actual pitch of aircraft, phase of pitch, phase lag of pitch, and z-scored classifier output *y* overlaid on joystick input onset. Potential feedback to the pilot is constructed by interpolating and filtering classifier output (cubic spline interpolation, then 3rd order Butterworth filter with 0.1 Hz cutoff to smooth output over 10s)

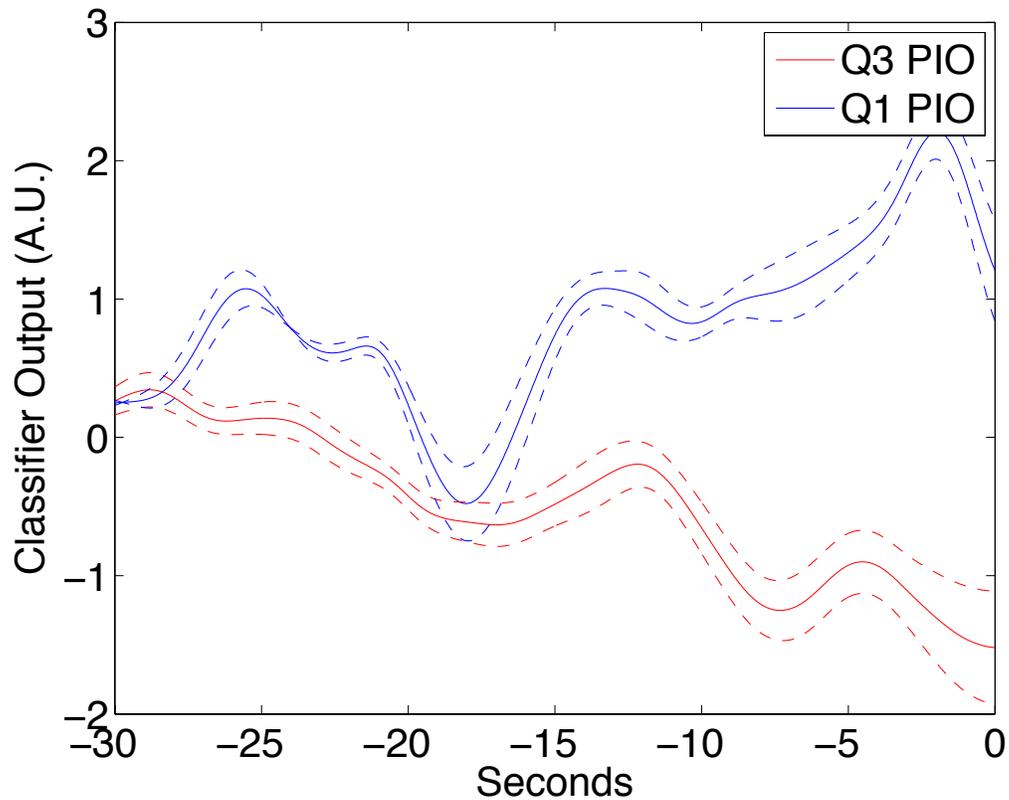

**Figure 8.** Figure shows that the output of workload classifier can be used to track PIO susceptibility in real-time. PIO susceptibility was estimated using a metric (0,1) based on joystick input (see Results). The last 5 seconds of each trial $i$ that ended in the medium sized ring (40-60s after first ring is crossed) were analyzed for PIO susceptibility; maximum value $M_i$ and time of maxima $T_i$ were computed. 'Q3 PIO' reflects the interpolated classifier output in the 30 seconds leading up to $T_i$ averaged over all trials where $M_i > 0.75$. 'Q1 PIO' reflects similar information for all trials where $M_i < 0.25$. This data suggests that even though the LvMS classifier only learns to differentiate EEG signals from different boundary conditions during trials, it can also differentiate piloting behavior. Note that data from -5s to 0s in the figure is guaranteed to reflect classifier output generated only within Medium glide path boundaries. Error-bars reflect mean ± SEM across subjects.

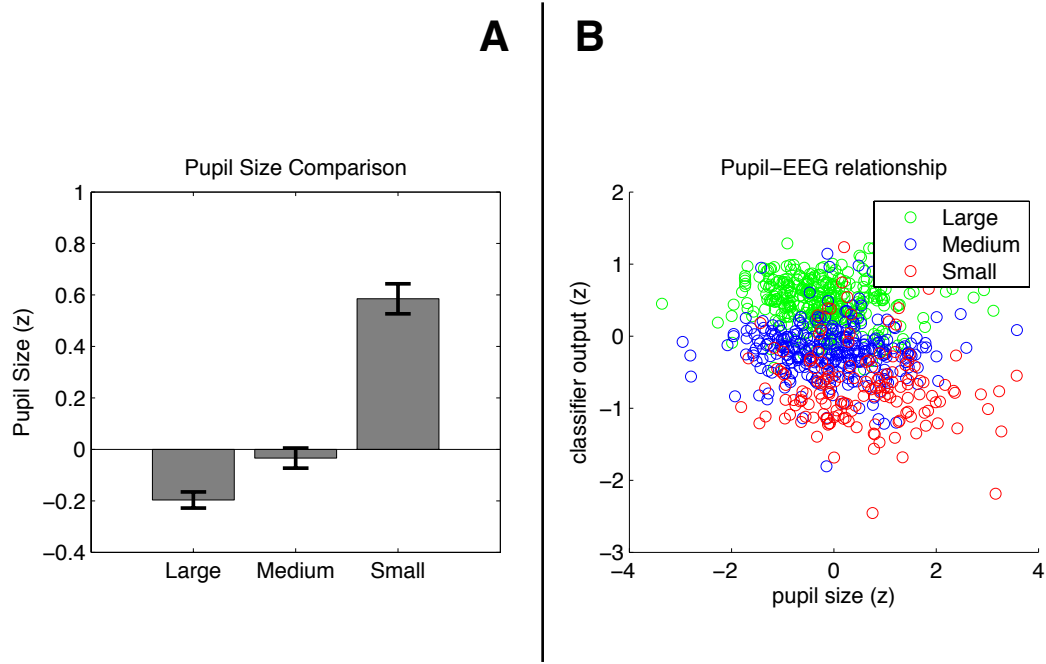

**Figure 9:** Pupillometry results. **A)** Data show an increase in pupil size (z-scored across each subject's data) with higher workload (smaller boundaries). **B)** The correlation between trial averaged classifier output (LvMS, full spectrum) and trial averaged pupil size, across all subjects (each datapoint represents a single trial). Overall correlation r=-0.1995, SEM=0.0764. 'Large' class correlation r = -0.0198, 'Small' class correlation r = -0.1210, difference n.s.

**A**

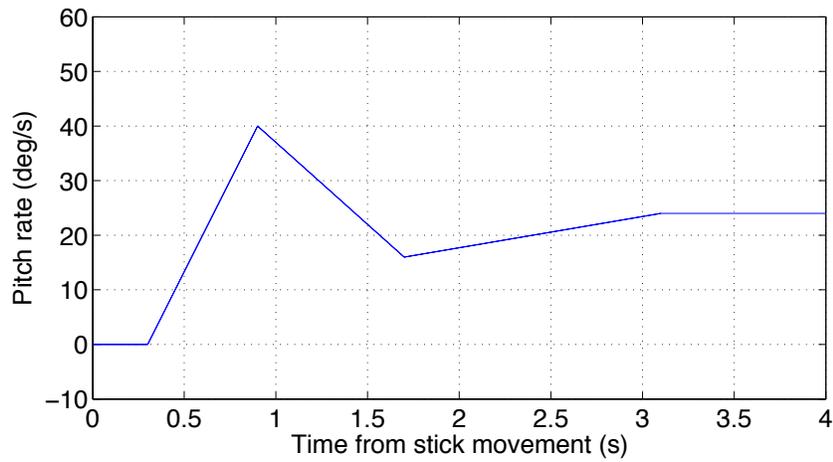

**B**

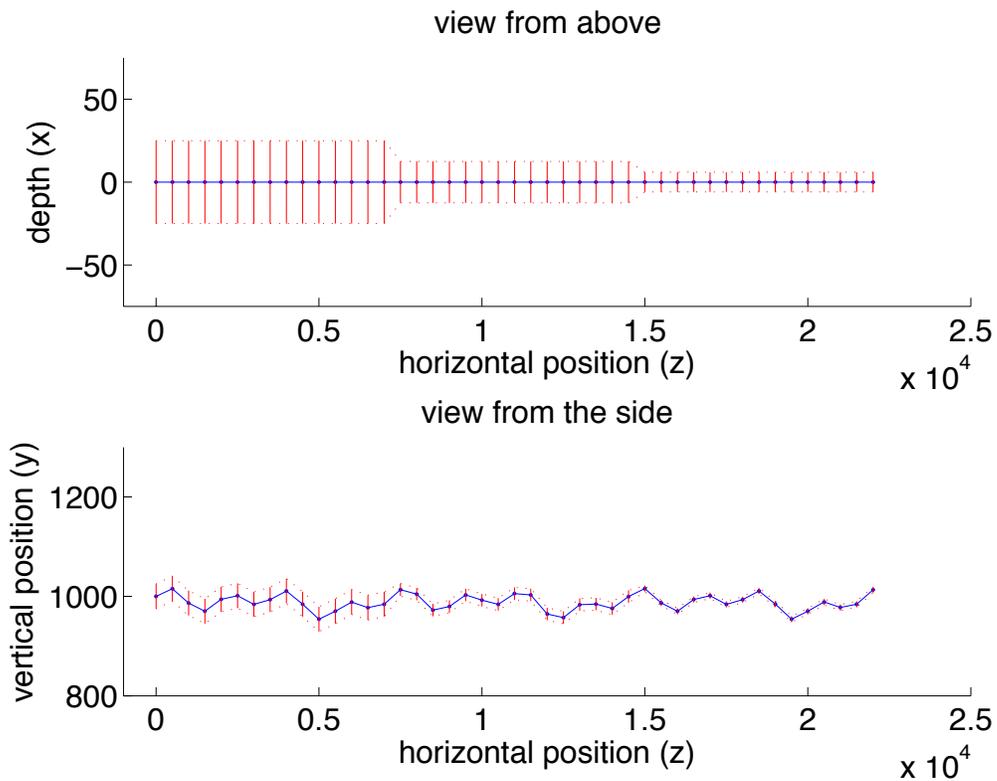

**Supplemental Figure 1. A)** Sustained pitch (y) response of the virtual aircraft to full joystick input (at time 0s on x-axis). The initial delay between input onset and response onset could either be 0.2s or 0.3s (0.3s in the figure). **B)** Full flight trajectory represented by the position and the size of glide boxes. In the view from the side, solid line shows the mean path through the center of glide boxes, while dotted lines describe glide box boundaries. All dimensions are in meters. Virtual aircraft moved steadily in z-axis @ 250 m/s and could be controlled in y (pitch) axis via joystick input. Half of all trials for each subject has this glide path trajectory, and the other half at trajectory shown in figure 1b.

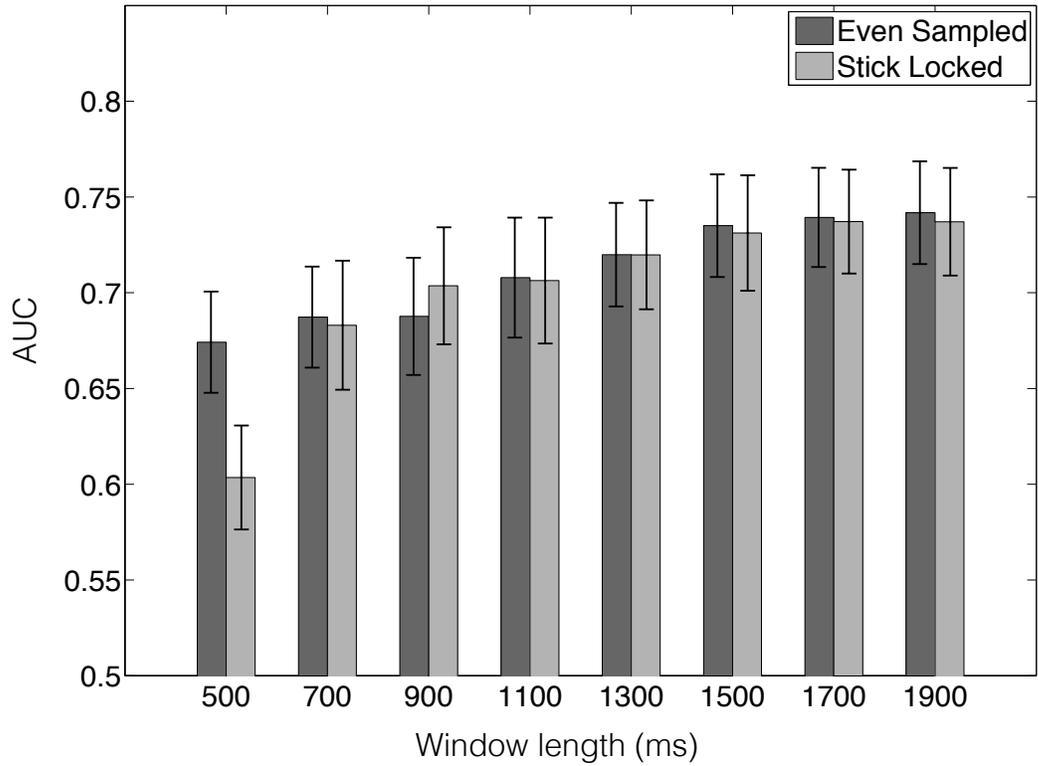

**Supplemental Figure 2.** The effect of increasing the length of EEG epochs, used to compute spectral features for each datapoint, on classification accuracy. AUC values asymptote around 1500ms, for both regularly sampled as well as stick locked classifier. Error-bars reflect SEM across 9 subjects.

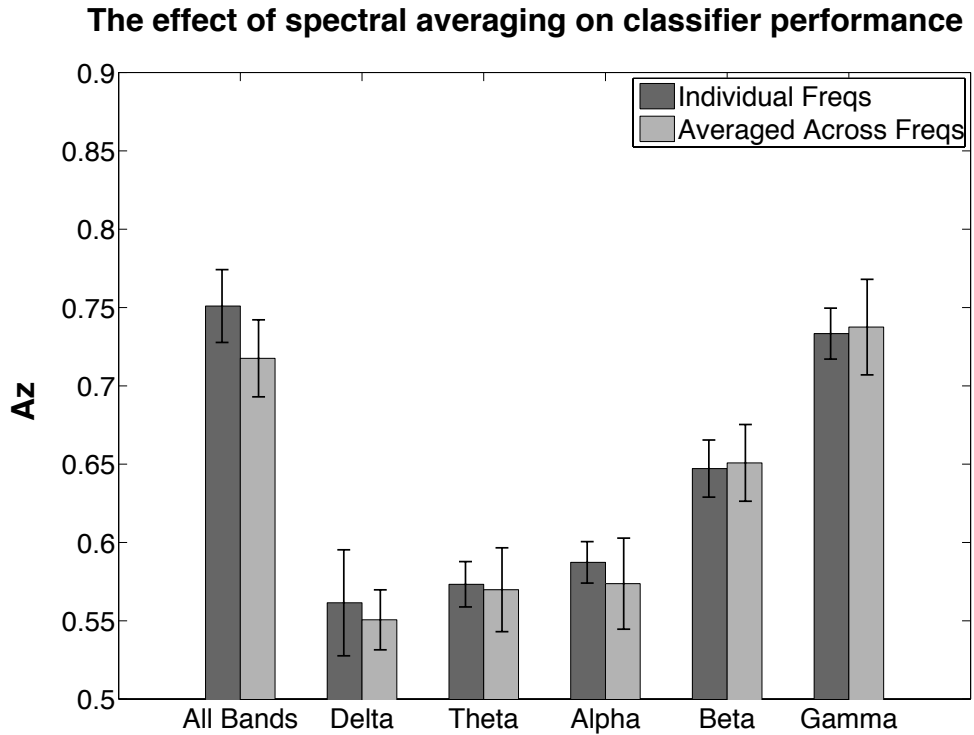

**Supplemental Figure 3.** Figure shows the comparison of AUC values for classifiers that used power of individual frequencies in each band as features versus classifiers that averaged power across frequencies in individual bands. The figure shows that the relatively higher contribution of Beta and Gamma band to overall classifier performance is not due to higher dimensionality of feature space when using individual frequencies as features. Error-bars reflect SEM across 9 subjects.

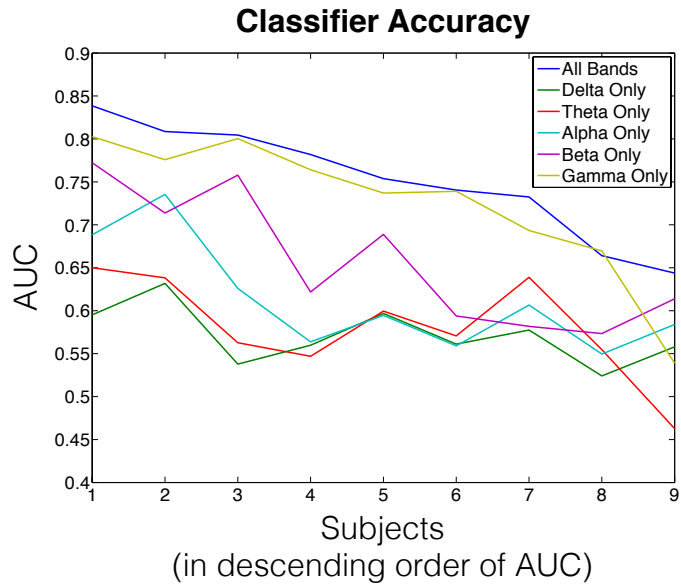

**A**

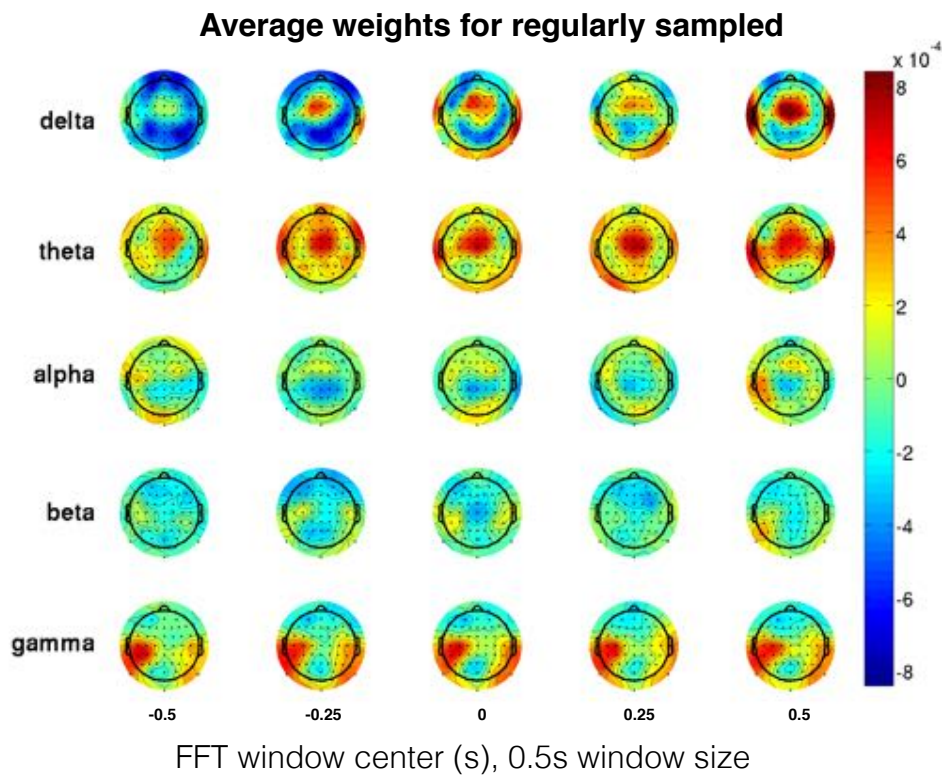

**B**

**Supplemental Figure 4. A)** AUC values for all subjects (in descending order of all band AUC), when using information from all spectral bands for regularly sampled data classification. Classification was performed using 64-channel MARA-cleaned EEG signals —1.5s epochs acquired every 2s — that were labeled according to the size of the nearest glide path boundary at the time of their generation. Also shown as AUC values for same subjects when using only individual bands (delta, theta, alpha, beta, or gamma). **B)** Subject-averaged scalp distribution of normalized weights for the classifying hyperplane that best separated Large boundary from Medium and Small boundaries (scalp map corresponds to 'All Bands' classifier in panel A).

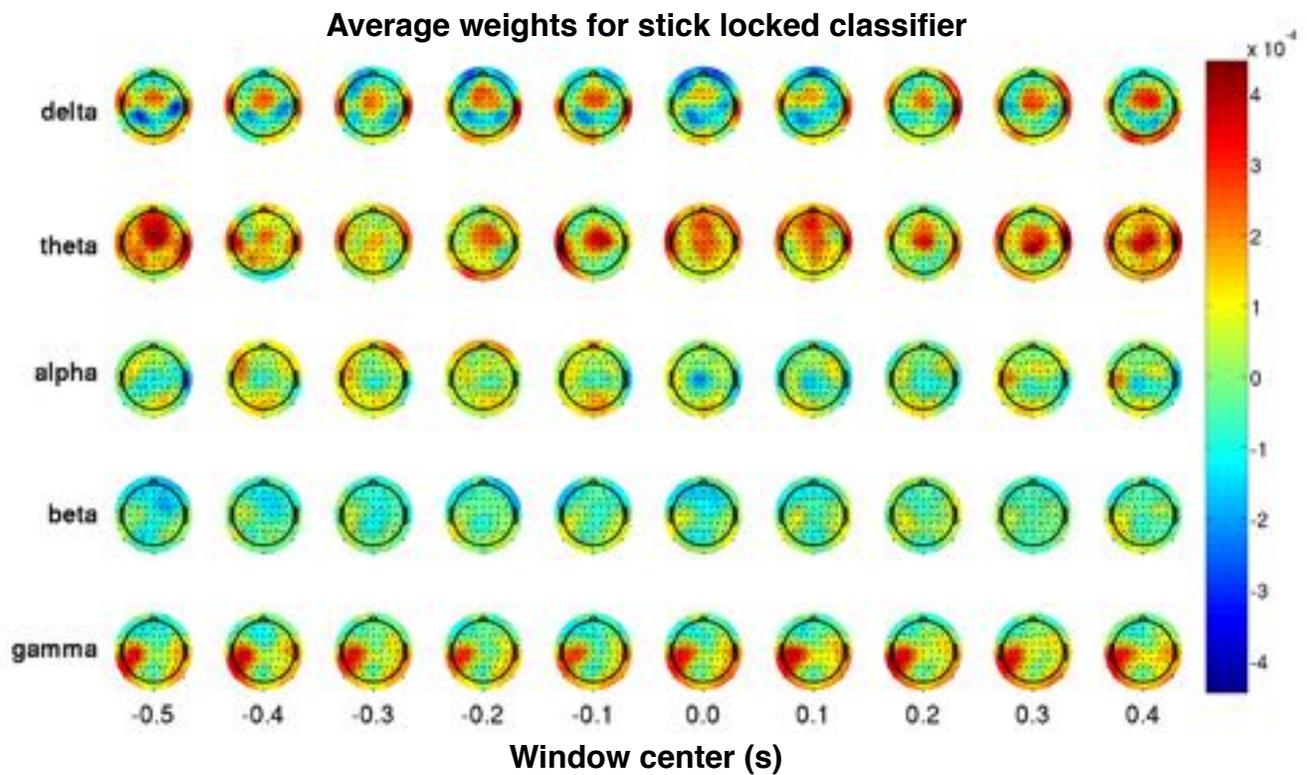

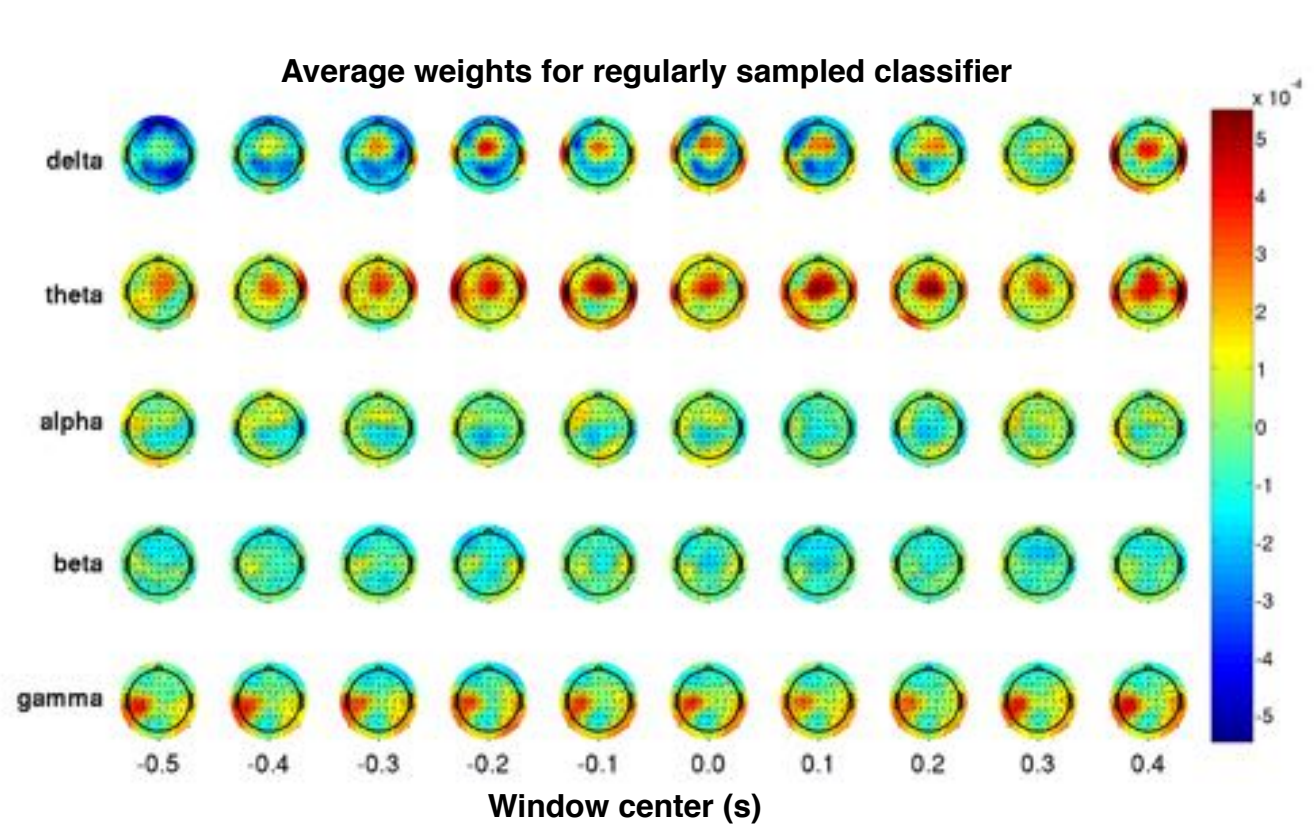

**Supplemental Figure 5. A)** Classification was performed using 64-channel 1.5s MARA-cleaned EEG epochs around each joystick movement, and were labeled according to the size of the nearest glide path boundary when they were generated. The Panel shows the scalp distribution of normalized weights for the classifying hyperplane that best separated Large boundary from Medium and Small boundaries. Delta and theta band activity from Fronto-central electrodes, as well as significant gamma band activity from a lateralized somatosensory topography, seems to contribute the most to classification accuracy. **B)** Figures show similar information as panel A, except that the classified EEG epochs that were acquired at regular intervals (2s).

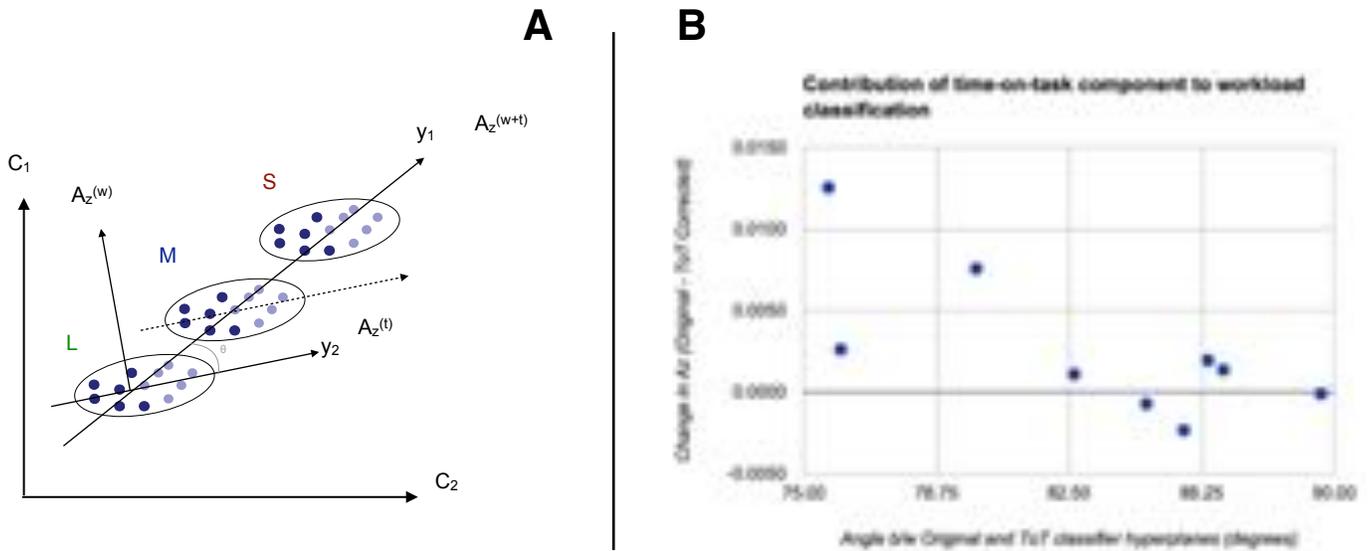

**Supplemental Figure 6.** Panel **A** shows schematic representation of hypothesized relationship of time-on-task (ToT) related signal to overall classification accuracy. Let C1 and C2 be two classifying features, e.g. power of a specific frequency at two electrodes. Let L, M, S be the data cloud generated while the subject operates under Large, Medium, and Small boundaries respectively. Dark colored points depict data sampled in the early half (first 15s) of flight within the respective boundary size, while light colored points depict the late half. Here, we assume that the classification accuracy of a workload classifier (L vs M/S) with AUC $A_z^{(w+t)}$ has a time-on-task component with AUC — $A_z^{(t)}$, i.e. a contribution of time since the start of the trial, that is unrelated to workload (or boundary change). Ideally, we would like to assess classification accuracy $A_z^{(w)}$ of pure workload component. Let the vector normal to the classifying hyperplane for a workload classifier (L vs M/S) be $y_1$, and the vector normal to the classifying hyperplane for a time-on-task classifier (L early vs L late) be $y_2$. Therefore, a large angle between $y_1$ and $y_2$ would imply a small contribution of time-on-task component to the accuracy of boundary based classifier for workload. Indeed, the scatterplot in Panel **B** shows that the angle between vectors normal to the hyperplanes for workload classifier and ToT classifier is quite large for each subject. Consequently, correcting for ToT component (through projection) produces negligible effect on AUC ($A_z$) values.